\documentclass[conference]{IEEEtran}

\ifCLASSINFOpdf
   \usepackage[pdftex]{graphicx}
\else
  \usepackage[dvips]{graphicx}
\fi
\usepackage{url}
\usepackage{subfigure}
\usepackage{cite}
\usepackage[]{fancyhdr} %
\newcommand{\changefont}{\fontsize{9}{9}\selectfont}
\IEEEoverridecommandlockouts
\begin{document}

%
\title{Modeling and Control of Multi-Energy Dynamical Systems: Hidden Paths to Decarbonization}

\author{\IEEEauthorblockN{ Marija Ili\'{c}\\Laboratory for Information and Decision Systems (LIDS)}
\IEEEauthorblockA{M.I.T.\\Cambridge, MA United States\\
ilic@mit.edu}
\and
\IEEEauthorblockN{Rupamathi Jaddivada\\Research Laboratory of Electronics (RLE)}
\IEEEauthorblockA{M.I.T\\Cambridge, MA United States\\
rjaddiva@mit.edu} 
\thanks{Partial funding provided by the US National Science Foundation }}


%





\maketitle
\thispagestyle{fancy}
\pagestyle{fancy}


\begin{abstract}
This paper points  out some key drawbacks of today's modeling and control underlying hierarchical electric power system operations and planning  as the hidden roadblocks on the way to decarbonization. We suggest that these can be overcome  by enhancing  today's information exchange and  control. This can be done by  revealing and utilising  inherent structure-preserving features  of complex  physical systems, and, based on this,  by establishing multi-layered energy modeling. Each module (component, control area, non-utility-owned entities) can be characterized in terms of its interaction variable, and higher level models can be used to understand   the interaction dynamics between different modules.  Once the structure is understood, we propose nonlinear energy control for these modules which supports feed-forward self-adaptation to ensure feasible interconnected system. Based on these technology agnostic structures it becomes possible to expand today's  Balancing Authorities (BA) to multi-layered interactive intelligent Balancing Authorities (iBAs) and to introduce protocols for flexible utilization of diverse technologies over broad ranges of temporal and spatial conditions.  
\end{abstract}



%
\IEEEpeerreviewmaketitle

\section{Introduction}
This paper concerns fundamental modeling and control of multi-energy dynamical systems  whose performance objectives are becoming rapidly diverse  and more complex than  there has been the case in the past.  As new investments are being planned, mainly for distributed energy resources (DERs) to replace large  polluting plants, it is necessary to  enable their on-line utilisation.  In Section \ref{sec:past}  we describe in some detail  some key power balancing and delivery problems facing the emerging electric power grid operations. In Section \ref{sec:new} we  propose basic enhancements necessary for enhanced operations and higher utilisation of resources  in a reliable and stable way.  In order to  manage  overly complex   and  system- and technology-specific interconnection rules  and approval processes, we  review  in Section \ref{sec:structure} inherent structures of today's electric power systems, and summarise in Section \ref{sec:newop}  structure-preserving modeling for managing complex systems with multiple administrative entities. We recall the notion of interaction variable introduced some time ago, and its fundamental role  in establishing multi-layered transparent models of different spatial and temporal granularity. As an example, we consider  small signal coordinated frequency stabilisation and regulation for systems with fluctuating intermittent resources in Section\ref{sec:smallfreq}. The rates at which these frequency oscillations occur are  basically electro-mechanical in nature, and, as such, can be modelled and controlled assuming no voltage dynamics.  However, we next recognise in Section \ref{sec:ibrint} that the increasing presence of inverter-based resources (IBRs) creates new problems such as very fast  electronically-induced  oscillations caused by the  interactions between the inverter controllers. 

The main contribution in this paper is presented in Section \ref{sec:general}  in which we propose a  unifying structure-preserving   multi-layered energy modeling of coupled electro-mechanical and electromagnetic dynamics, and its inherent structure-preserving properties. A modular aggregate model for representing energy system  dynamics is relevant for assessing and controlling  interactions between subsystems, and it is based on   a recently introduced generalised interaction variable as a means of specifying the   ability of subsystem to inject power at certain rates.  Notably, we stress the   key role  of rate of change of instantaneous reactive power \cite{WyattIlic}, its modeling and the validity for general non-sinusoidal voltages and currents.  The second  major contribution is described in Section \ref{sec:control}:  Using multi-layered energy modeling  we  conceptualise a  unifying technology-agnostic energy control of interactions between the components which results in a linear closed-loop aggregate energy interactive model. This nonlinear energy control  has a transparent and explainable  physical meaning, and 
it lends itself to  provable performance of large-scale  system dynamics  over broad ranges of input and topology changes. This is particularly important during extreme events, faults  and loss of large generation in bulk power systems.  Similar control is needed in  small microgrids subject to wind gusts and large  weather-related  variations in solar  radiance. Since the control  does not require real-reactive power decoupling nor linearisation it becomes effective during large changes of operating conditions, an extremely difficult task.  We suggest   three steps toward generalising today's area control error (ACE) as the basis for  reliable and efficient  emerging  systems operation.  
\section{Fundamental challenges to today's hierarchical modeling and control}
\label{sec:past}
As new resources are  beginning to replace large-scale controllable power plants, the industry is facing many major operations and planning challenges. In deregulated industry  many mandates require utilities to connect new resources in an open-access manner, and the interconnection process is time consuming. Even in the regulated industry it  has become difficult to use  today's worst case scenario reliability and security tests  in  systems with highly intermittent resources.  More generally, today's deterministic rules  do not lend themselves well to operating  under uncertainties  \cite{perform}.   Important for purposes of this paper is that  the changing systems are beginning to experience  dynamical problems, such as  low  frequency oscillations  \cite{nordic} as well  as  fast control-induced sub synchronous stability (CISS) problems \cite{ercot,australia,hawaii,germany}. 
This situation  brings up many technical challenges of interest in this paper, in addition to regulatory and financial. Replacing  large polluting power plants with  smaller-scale distributed  intermittent  resources requires new investment in   grid   wires  as well as in non-wire  solutions. Non-wire solutions are mainly ``smarts"  including various types of FACTS, HVDC, and controllers placed on both conventional power plants (governors, AVRs, PSSs)  and inverter-based resources (IBRs), which comprise power-electronically controlled solar  and wind power plants, batteries and even clusters of EVs.   It quickly becomes obvious that the emerging systems are going to have many more primary  controllers dispersed throughout the system, and their design and tuning represents a major technical challenge, summarized next. 
\section{Necessary enhancements of today's operations}
\label{sec:new}
Advanced bulk power systems (BPS) are operated in a hierarchical manner,  by performing feed-forward power scheduling to supply predictable system load (tertiary control);  frequency (and in some systems voltage)  is regulated in  response to relatively slow deviations from schedules by the Balancing Authorities (BAs) (secondary control); power plants  have local primary controllers, governors, AVRs and some times PSSs,  and are expected to stabilize fast voltage and frequency fluctuations to their set values given by the higher level controllers \cite{shell,iliczab}. 

However, today's hierarchical control needs to be enhanced  to be effective in the emerging systems with  high penetration of intermittent resources.  Tertiary level  computer applications need to be such that they  enable participation  by   all controllable resources, and not only  by large power plants, as it is mainly the case today. Also, it is necessary to schedule all controllable variables (voltage set points, adjustable demand) in addition to real power dispatch. This is needed because it is often necessary to  optimize voltage dispatch in  order to support power delivery over large electrical distances.  While voltage is often thought of  as a highly  local problem, computer tools like AC Optimal Power Flow (AC OPF) are needed to find  voltage  binding constraints which make the AC power flow infeasible and to advise  which voltage and other adjustments are needed to make the AC power flow feasible within acceptable operating constraints  \cite{netssacopf}. Also, tertiary level scheduling tools should  have adaptive performance metrics to seamlessly find a combination of actions (real power, voltage,  relaxing line flow limits,  load shedding) so that a feasible operation is found for anticipated uncontrolled system demand. These functions are critical  in systems where power flow analyses and/or real power generation dispatch are insufficient to  enable feasibility. Examples of these are striking, it was shown that 1 GW more hydro power can be delivered from Niagara Mohawk to NYC on a hot summer day by combining voltage and real power dispatch \cite{nyserda}.  These functions are critically important during $(N-k)$  where $k>>2$ contingencies typical of weather-related extreme events such as hurricanes in Puerto Rico  \cite{puertorico} and, more recently, in ERCOT \cite{ercoth}.  Having such resource allocation tools  is very important for managing systems with high penetration of intermittent resources as  they support less conservative preventive and corrective, flexible reserves and ensure their delivery, namely  feasible AC power flow.  Finally,  tertiary level SCADA should be expanded to interactively exchange information with lower level grid users whose ability  and willingness to participate in grid services may vary. Much the same way, as conventional power plants have ramp rate-limited ability, the less conventional resources can and should be integrated into tertiary level by relying on their information about power and rate of change of power.  At the tertiary level this information  is needed in a feed-forward manner, prior to scheduling. A seemingly minor, but important is the differentiation between the system operator using ramp-rate limit as a hard constraint when scheduling, on one side, and optimising use of diverse resources within power and rate of change of power ranges determined by the resources themselves in a  model predictive manner, on the other  \cite{mpc}.   

\begin{figure}[!htbp]
\centering
\includegraphics[width=0.95\linewidth]{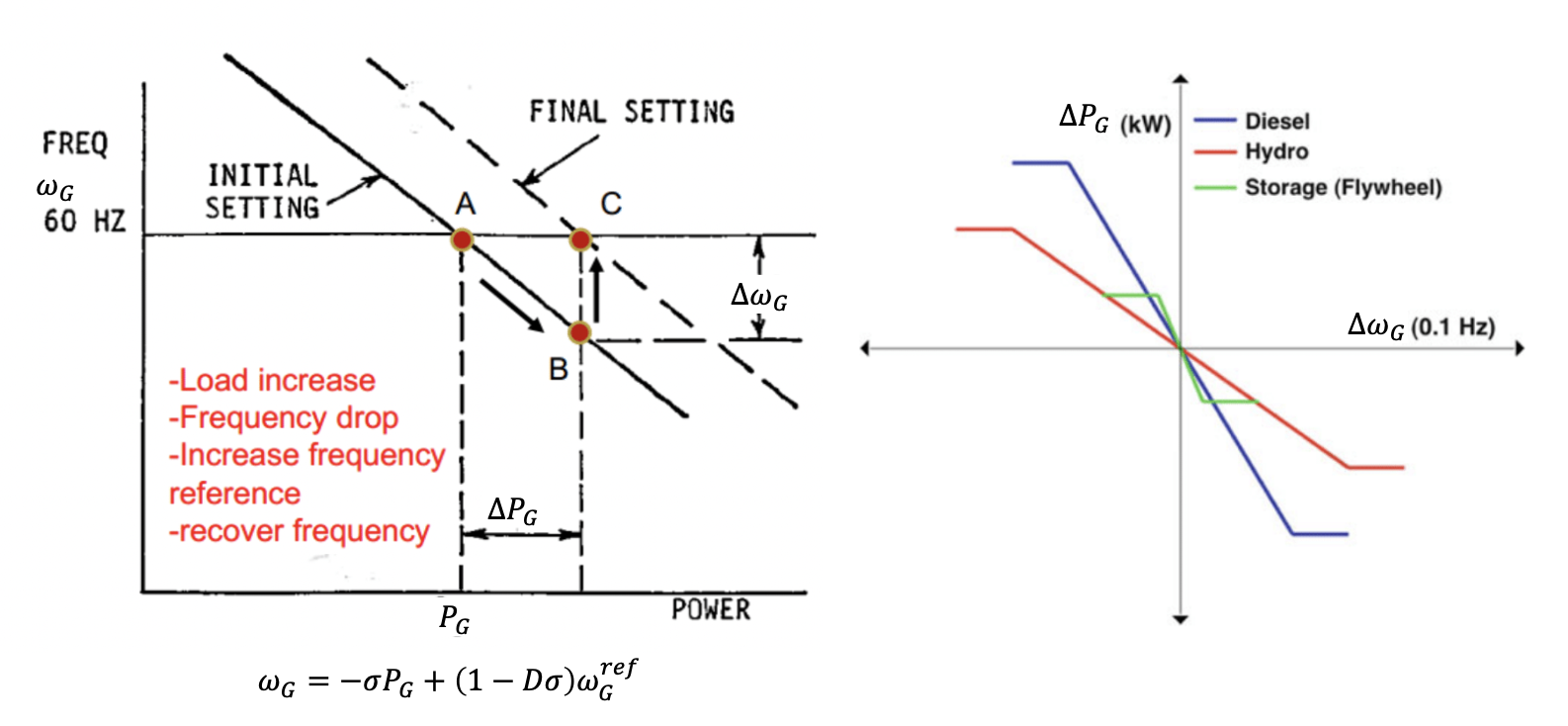}
\caption{Basic function of secondary control: Adjusting reference set point to produce given power,assuming nominal frequency}
\label{fig:droop2}
\end{figure}

\begin{figure*}[!htbp]
\centering
\includegraphics [width=0.95\linewidth]{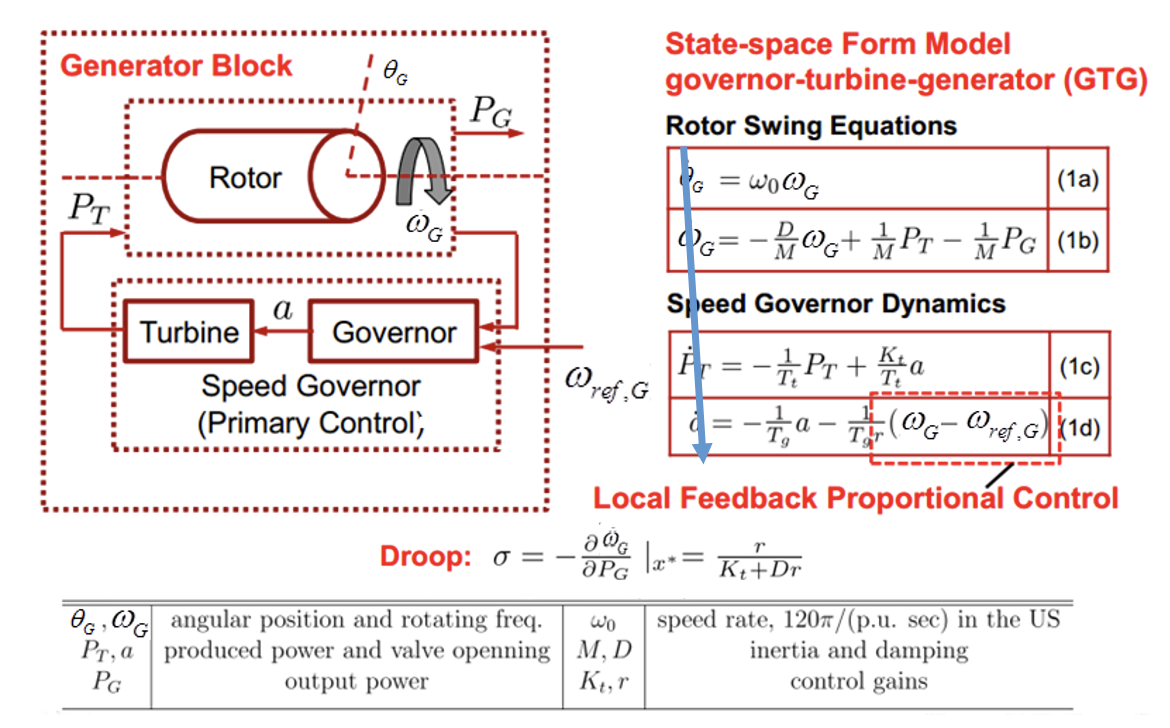}
\caption{Sketch of a generator-turbine-governor (G-T-G) and its droop model}
\label{GTGset}
\end{figure*}

The actual implementation of these commands is carried out by the governor changing the set point for its  ``droop" $P_G^{ref}[K]$ in a feed-forward manner ($K$ represents the timescales relevant for tertiary control). Secondary level control is generally used to map  power commands given by the tertiary level into the set points  to which controllable equipment responds and/or to in a feedback manner adjust frequency set points so that  incremental power is produced and the system frequency deviation within a pre-specified tolerance of $\epsilon$ is allowed. However, 
the line between tertiary and secondary  level governor control is often gray. In systems which do not have automatic generation control  (AGC), only ``power sharing" commands are implemented this way.   Regulation  of frequency deviations resulting from relatively slow  power excursions around the scheduled value is done manually in such systems. In systems with AGC, the PI governor control responding to frequency deviations, and/or area control error $(ACE)$  in multi-BA systems, adjusts the set point for governor controller $\omega_G^{ref} [k]$ ($k$ denotes discrete time samples relevant for secondary control) according to the three way  droop relation. Shown in Figure  \ref{fig:droop2}  is a sketch of these droops defining relations between frequency deviations from nominal frequency $\omega_0$,  $\Delta \omega_G[k] = \omega_G[k] - \omega_0$   and power increments $\Delta P_G [k] = P_G [k] -  P_G^{ref}[K]$ for given  $\omega_G^{ref} [k]$. AGC basically computes set point $\Delta \omega_G^{ref}[k] = \omega_G^{ref}[k+1] - \omega_G^{ref} [k] $  so that $\Delta \omega_G[k]  < \epsilon$ \cite{iliczab}.  To avoid a major confusion in recent literature concerning the notion of a droop in systems with IBRs, in particular, we point out that  power sharing takes place in a feed forward way for predictable bounds on power deviations, while the AGC takes place in a feedback manner at the faster rate. Moreover, the three way droop  used for AGC is a quasi-static concept which is derived by  assuming that derivatives   
are zero as illustrated in Figure \ref{GTGset}. The power sharing droop can be set independently  in a slower feed forward way to shape the frequency response to anticipated power by the tertiary level.  This clarification is written because  many references in the literature assume one or the other  functionality of secondary level control. 
There are several  major issues regarding the role of secondary level control. The main is that power sharing function assumes that governor can control power generated over broad ranges of  power commands given by the tertiary level.  The second major issue is that the gains in governor controller are derived using a linearised turbine-generator model around pre-selected operating point. The third major issue is that secondary level assumes that primary controller stabilises and regulates output variables (frequency) to its set point, and, that it can control power generated  at the right rate.  All these assumptions are major causes of frequency problems in systems with intermittent resources and require further studies as discussed in Section \ref{sec:structure}. 

Finally,  and independent from the version of secondary control in place, a  combination of primary and secondary control should guarantee that commands given by  the tertiary level  are implementable, namely stable, feasible and robust with respect to parameter uncertainties. In this paper we are particularly concerned with the huge issue  that  state-of-the-art of today's primary controllers can generally not do this, and propose new controllers.  Broadly speaking, they can not stabilise system dynamics in response to continuously varying  fast power imbalances. It is generally  hard to control power/rate of change of power while maintaining voltage and frequency within the operating limits. Later part of this paper describes this in some detail. 

\section{Structure-preserving modeling of  frequency dynamics  in transformed state space}
\label{sec:structure}
\begin{figure}[!htbp]
\centering
\includegraphics[width = 0.9\linewidth]{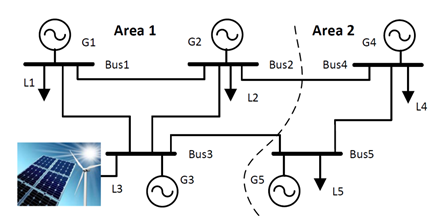}
\caption{A sketch of two-area interconnected power system}
\label{fig:twoarea}
\end{figure}

In order to  overcome extreme perceived complexity  when attempting to  improve today's hierarchical control  so that it supports the  on-going changes,  we propose to consider the structure-preserving modeling of electro-mechanical dynamics relevant for assessing frequency stability, briefly summarised next. Consider without loss of generality a two-area electric power system shown in Figure \ref{fig:twoarea} with highly varying solar power in one of the areas. 
It is known that the linearized electro-mechanical dynamics 
$x_{LC,i}$
of  each module  $i$   ( resource, load, BA, interconnected system)  can be written in terms of its  internal local state variables $x_{LC,i}$, 
and the real power exchanged with the rest of the system $P_{G,i}$.  As an illustration, shown in Figure \ref{GTGset} is a conventional generator with its governor control. Its states $x_{LC,i}$ are shown in  Figure \ref{GTGset}, and their dynamics are  
\begin{eqnarray}
{{\dot x}_{LC,i}}
= \underbrace {\left[ {\begin{array}{*{20}{c}}
0&{{\omega _0}}&0&0\\
0&{\frac{{{K_d}}}{M}}&{\frac{1}{M}}&0\\
0&0&{ - \frac{1}{{{T_t}}}}&{\frac{{{K_t}}}{{{T_t}}}}\\
0&0&0&{ - \frac{r}{{{T_g}}}}
\end{array}} \right]}_{{A_{LC}}}{x_{LC,i}} + \underbrace {\left[ {\begin{array}{*{20}{c}}
0\\
{\frac{1}{M}}\\
{\;0}\\
0
\end{array}} \right]}_{{c_M}}{P_{G,i}}
\label{eqn:gendyn}
\end{eqnarray}

It can be seen that this model takes on the standard state space form in which local state variables  $ x_{LC,i} =[\theta_{G,i} ~~\omega_{G.i} ~~ P_{T,i}~~a_i]^T$  define the dynamics as a function of  power
generated $P_{G,i}$; notably they do not depend on $\theta_{G,i}$  explicitly \cite{naps94}. Power generated and its rate of change  must balance local load deviation and any other local disturbance as follows
\begin{equation}
\dot  P_{G,i} =  \dot P_{L,i} 
\label{eqn:ratepower}
\end{equation}
When Equations (\ref{eqn:gendyn}) and (\ref{eqn:ratepower}) are combined they result in dynamical model in transformed state space of a stand alone   generator $i$ 
\begin{eqnarray}
\underbrace{\left[ {\begin{array}{*{20}{c}}
{{{\dot x}_{LC,i}}}\\
{{{\dot P}_{G,i}}}
\end{array}} \right]}_{\dot{x}_i^{new}} = \underbrace{\left[ {\begin{array}{*{20}{c}}
{{A_{LC}}}&{{C_M}}\\
{{K_P}}&0
\end{array}} \right]}_{A_i}x_i^{new} + \underbrace{ \left[ {\begin{array}{*{20}{c}}
0\\
1
\end{array}} \right]}_{D_i} {{{\dot P}_{L,i}}}
  \label{eqn:genmodel}
\end{eqnarray}
  
where 
\begin{equation}
x_i^{new} = [\omega_G ~~ P_T~~a ~~P_G]^T 
\label{eqn:newstatespace}
\end{equation}
is the new state variable in transformed state space. 

Consider now a  BA $J, J\in \{1,2\}$  of two area system shown in Figure \ref{fig:twoarea}.  When generators  are interconnected, the model of their local states Eqn. (\ref{eqn:gendyn})  remains the same as for the stand-alone generator case  and takes on the form
\begin{equation}
    \dot{x}_{LC} = A_{LC} x_{LC} + C_M P_G 
    \label{eqn:allgendyn}
\end{equation}
and the conservation of power becomes
\begin{equation}
 \dot P_G =K_P \omega_G  +\dot F_e + D_P \dot P_L  
 \label{eqn:areaconstraints}
\end{equation}
where $P_G$, $\omega_G$, $F_e$ and $P_L$  are vectors of power generated by all  generators in the area, net tie-line power flow coming into the area and loads within the area \cite{naps94}.  Matrix $K_P$ is structurally singular as a direct result of conservation of energy in the area. 
Equations (\ref{eqn:allgendyn}) and (\ref{eqn:areaconstraints}) combinedly result in a decentralised model of BA in transformed state space \cite{shell} 
\begin{eqnarray}
\left[ {\begin{array}{*{20}{c}}
{{{\dot x}_{LC}}}\\
{{{\dot P}_G}}
\end{array}} \right] = \left[ {\begin{array}{*{20}{c}}
{{A_{LC}}}&{{C_M}}\\
{{K_P}}&0
\end{array}} \right]\left[ {\begin{array}{*{20}{c}}
{{x_{LC}}}\\
{{P_G}}
\end{array}} \right] + \left[ {\begin{array}{*{20}{c}}
0\\
{{{\dot F}_e}}
\end{array}} \right] + \left[ {\begin{array}{*{20}{c}}
0\\
{{{\dot P}_L}}
\end{array}} \right]
\label{eqn:areamodel}
\end{eqnarray}

Once this structure is understood, it becomes fairly straightforward to formalize objectives of primary and secondary level generation control to both stabilize and regulate frequency in systems whose disturbances are continuously varying loads and/or intermittent resources. We observe that today's separation of primary control for frequency stabilization, on one side, and secondary control for frequency regulation, on the other, can no longer be done by assuming that secondary control is quasi-static and that, as such, can be modeled using quasi-static frequency regulation  droop-based  models and by neglecting the grid effects \cite{automatica,kundur}. Instead, continuous-time  models  are needed to stabilize and regulate frequency in systems with  ever-fluctuating power imbalances \cite{chow}. 
\begin{figure}
\centering
\includegraphics[scale=.6]{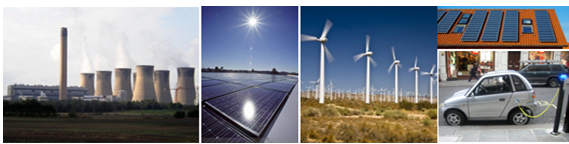}
\caption{Emerging diverse technologies}
\label{fig:allcomp}
\end{figure}
Second, quite relevant for the changing systems is that the same structure-preserving modeling is technology agnostic; shown in Figure \ref{fig:allcomp} are vastly different technologies, all of which lend themselves to the same structural modeling. 

\section{Structure-preserving modeling of systems comprising multiple entities}
\label{sec:newop}
\begin{figure}[!htbp]
\centering
\includegraphics[width=2.5in]{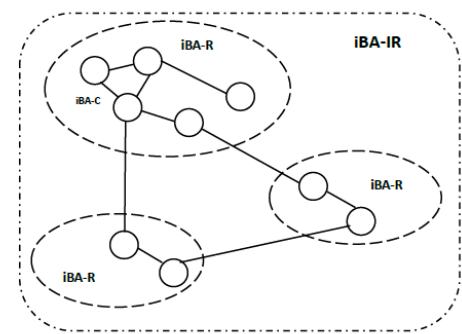}
\caption{Emerging portfolios of  intelligent Balancing Authorities (iBAs)}
\label{fig:nestediBAs}
\end{figure}
\begin{figure}[!htbp]
\centering
\includegraphics[width=2.5in]{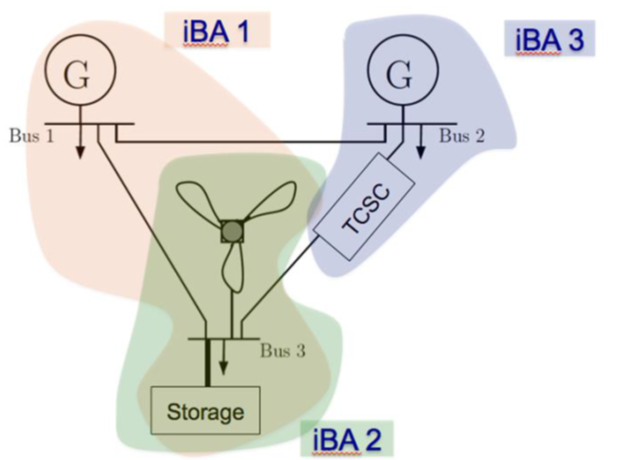}
\caption{Emerging portfolios of  intelligent Balancing Authorities (iBAs)}
\label{fig:portfolia}
\end{figure}

To further formalise  modeling  in the changing industry, we suggest that one of the major challenges is the lack of specifications of new technologies in terms of variables that make it possible to integrate them the same way as other resources.  Shown in Figure \ref{fig:nestediBAs} is a sketch of the emerging architectures, in which often non-utility-owned entities wish to connect to the legacy system comprising coarse BAs \cite{iBAs}.  These new entities are ``nested" within the legacy BAs, and often have their own control and decision-making ``smarts".  Such emerging examples are: portfolios of wind farms and storage (See Figure \ref{fig:portfolia}); interconnected microgrids embedded in today's distribution grids with and/or without own storage; clusters of EVs; neighbourhoods with controlled  water pumps,  HVACs, and water heaters; clusters of smart buildings;  utility scale solar plants;  and many more. 

It is highly unrealistic to require that each of these new technologies  reveals exactly their internal dynamical models and control. To overcome this problem, we propose that these enities can be viewed as technology-agnostic iBAs  which can be characterized in terms of their interaction variables (intVar). 
This notion of intVar was introduced first  for linearized systems  for frequency stabilization and regulation and has  two unique properties key to modeling, assessing and controlling interactions between different entities. In short, an intVar is a direct consequence of energy conservation  by each iBA and, when modeling only linearized electro-mechanicial  interactions can be shown to  have the following \cite{shell,iliczab}:
\begin{itemize}
    \item Property 1: intVar $z_i$  associated with component $i$ is a function of its own internal  state variables written in transformed power state space, Eqn. (\ref{eqn:areamodel})
    \item Property 2: When disconnected from the rest of the system  it remains constant $\frac{d z_i}{dt}=0$. 
\end{itemize}
The existence of interaction variable associated with any physical  module (generator $Gi$,  BA $J$) whose structure-based model in transformed state space was introduced above directly follows from the fact that its  system matrix is structurally singular. For the case of BA $J$ $K_P$ in its transformed state space model (\ref{eqn:areamodel}) matrix $K_P$ is structurally singular, leading to structurally singular matrix in its transfomred state space given in Eqn. (\ref{eqn:areamodel}).

Therefore, there exists a transformation  $T_{BA}$ 
such that 
\begin{equation}
    T_{BA} A_{BA} =0 
\end{equation}
where $A_{BA}$ is system  matrix in  BA model  (\ref{eqn:areamodel}).  Then, by pre-multiplying this model by such transformation 
it follows that
\begin{equation}
\dot z_{BA} = T_{BA} \dot F_e + T_{BA}D_P \dot P_L
\label{eqn:intvarBA}
\end{equation}
The  interaction variable is defined as
\begin{equation}
z_{BA}=T_{BA} x_{BA}^{new}
\label{eqn:intBa}
\end{equation}
is the integral of power imbalance created by deviations in net tie line flow power and internal load power deviations.  Similarly, it can be shown that  $T_{Gi}$ exists such that 
\begin{equation}
    T_{Gi} A_{Gi} = 0
    \label{eqn:Tgeni}
\end{equation}
where $A_{Gi}$ is the  matrix in the generator model  given  in Equation (\ref{eqn:gendyn}). Its  interaction variable can be shown to be
\begin{equation}
    z_{Gi} = T_{Gi} x_{Gi}^{new}
    \label{eqn:intgeni}
\end{equation}

The interaction variable has a straightforward physical interpretation. It is directly a result of first  conservation law of energy  which says that the dynamics of stored energy $E_i$  in a module $i$ is a result of net power injected into it  $P_i$ minus its  internal thermal loss $\frac{E_i}{\tau_i}$, as follows
\begin{equation}
    \dot E_i = -\frac{E_i}{\tau_i} + \dot z_i 
    \label{eqn:intvari}
\end{equation}
In the case of linearized electro-mechanical dynamics $\dot z_i =P_i$, where $P_i$ is real power out of module $i$ and $\tau_i$ is time constant representing thermal  losses. 
\begin{figure}[!htbp]
\centering
\includegraphics[width=0.9\linewidth]{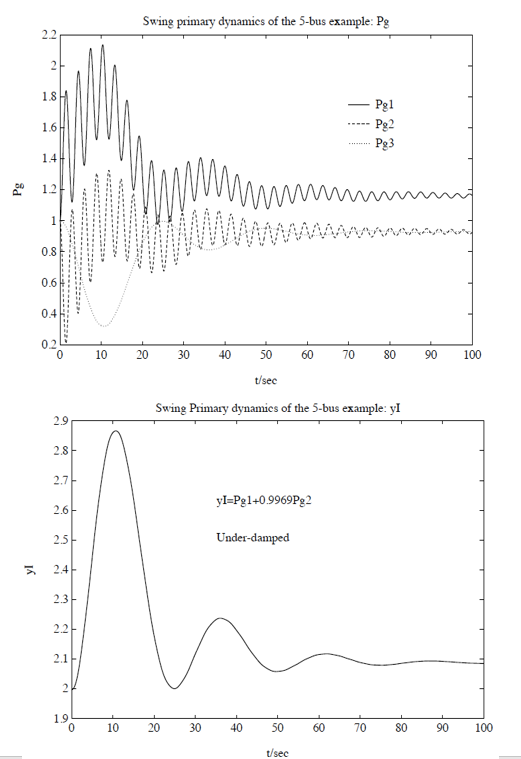}
\caption{Interaction variables and power generated by individual plants}
\label{fig:intvarsZ}
\end{figure}
Shown in Figure \ref{fig:intvarsZ} are interaction variables and internal generator  power variables  for the two area system shown in Figure \ref{fig:twoarea}  when solar power radiance is zero. Only electro-mechanical  dynamics can be   modeled for assessing inter-area power  and frequency oscillations. We  generalize the notion of intVars in Section \ref{sec:ibrint} that can capture the electro-magnetic fast oscillations  and the effects of their inverter cotnrol. 

\section{Structure-based small signal frequency stability assessment}
\label{sec:smallfreq}
Representing small signal frequency dynamics of very large-scale multi-BA with nested iBA systems as in Figure \ref{fig:nestediBAs},  is almost an impossible task in the changing industry. The resources and loads can have very different manufacturers data, and include many diverse types,  hydro, coal, gas, wind and their diverse non-standardized  governor controllers.  Also,  many  entities do not like to share their internal technology. To overcome this problem, we  propose here that sources of low frequency oscillations caused by  poor tuning of governors, for example,  can be  detected in a highly structure-preserving transparent manner by exchanging only  minimal information  in terms of their intVars, a method based on \cite{qixingphd}. 
The method is based on conceptualising the system shown in Figure \ref{fig:twoarea}  as a large-scale interconnected system comprising two subsystems, shown in Figure \ref{fig:twoarea}.  A vector Lyapunov function  $v(x) = [v_1(x_1) ~ v_2 (x_2)]^T$ has  Lyapunov functions of each subsystem $v_1(x_1) = x_1H_1x_1^T$ and $v_2(x_2) = x_2H_2x_2^T$. Matrices $H_1$ and $H_2$  are  computed by solving Riccati equations in each subsystem
\begin{eqnarray}
A_1H_1 + A_1^TH_1 & = & -G_1\\
A_2H_2 + A_2^TH_2 & = & -G_2
\end{eqnarray}
where $G_1$ and $G_2$ are positive definite matrices.  It can be shown that the vector Lyapunov function has negative derivative if matrix $W$ with its entries defined as 
\begin{eqnarray}
w_{ij} & = & - \frac{1}{2} \frac{\lambda_m(G_i)}{\lambda_M (H_i)} + \lambda_M^{\frac{1}{2}}(A_{ij}^T A_{ij})  ~~ i = j \\
w_{ij} & =& \lambda_M^{\frac{1}{2}}(A_{ij}^T A_{ij}) ~~ i \ne j
\end{eqnarray}
is a Metzler matrix and it is  diagonally dominant \cite{siljak}. Here $\lambda_M$ and $\lambda_m$ are maximum and minimum eigenvalues, respectively. 

To check this condition, each subsystem computes in a decentralized manner the minimal and maximum eigenvalues and tests whether  the interconnected system is made unstable because of very strong effects of interconnection matrices $A_{ij}$.  This method is computationally very effective because it only requires computing extreme eigenvalues by the subsystems themselves, and checking conditions of the  very low order matrix $W$ whose dimension is  determined by the number of subsystems.  This method, in addition to  being scalable,  indicates relative effects of local stability and interactions on global stability of the interconnected system. 
The same method can be used to design control of subsystems as well as the control of FACTS devices which can affect the strength of interactions on tie lines interconnecting subsystems.  For numerical examples  illustrating  use of these conditions on  the two area system shown in Figure \ref{fig:twoarea}, see \cite{qixingphd}. 
\section{Examples of IBR-and fault-caused electromagnetic voltage  oscillations}
\label{sec:ibrint}
The electro-magnetic voltage  oscillations have  typically been  a problem during  sudden fast faults in BPS.  However,  they are also  becoming  an important challenge in systems with fast fluctuating intermittent resources, due to, wind gusts and hard-to-predict variations in solar radiance. There has been a major concern about ``loss of inertia" and the need  for faster-responding controllers to low frequency voltage oscillations  in such systems. At present there is a major effort under way to   control wind and solar power fluctuations with their own fast  power-electronically-switched  controllers and these resources are referred to as the IBRs  \cite{TFreport,NREL}. There are also many power-electronically switched controllers placed in grid components, such as  HVDC technologies,  STATCOMs, Static Var Compensators (SVCs), Thyristor Controlled Series Capacitors  (TCSCs) and, more recently, even electronically controlled series inductors \cite{smartwires}.   

While system planners continue to analyse likely operating problems in these newly emerging systems by   relying on well-established 
feasibility, small signal stability and transient stability  analysis tools, it is becoming increasingly clear that both enhanced modeling and new controllers are needed.  The  currently used software  often does not even  identify  transient instability, and, as a result, can not assess potential benefits of deploying fast switching controllers. The industry is aware of these issues and it often resorts to  using  a very  detailed electromagnetic transients (EMT)  modeling  to set and test protection  during faults, for example. These tests are excessively time consuming and system-specific. Because of their complexity, they  are not scalable and, most critically,  can not be used to assess  interactions between an IBR tested and other dynamically varying  IBRs and grid controllers placed elsewhere in the system. The tests are typically  done using static Thevenin equivalent of the rest of the system and applying a combination of most likely and/or most critical step   power changes at the IBR location \cite{tests}. System measurements detect interactions between, for example, an  IBR wind power plant and connecting weak TCSC-compensated  transmission line, but there are no effective software tools for simulation of this ``fighting" of IBR controllers known as control induced sub-synchronous stability (CISS) problems   \cite{ercot}. 
In this paper we propose that these  oscillations caused by the IBRs and the smarts on the grid interacting can be modeled to a large extent using structure-preserving modeling  described earlier in this paper   by introducing a  generalised interaction variable. In addition to modeling electro-mechanical stored energy the generalised energy modeling  represents fast electro-magnetic dynamics as well. This is introduced in the next Section  \ref{sec:general}.   In this section we briefly illustrate   the electromagnetic energy and its dependence on the controller logic used. Notably, these examples utilize TVP  modeling of all components, including line currents and voltages and do not require detailed EMT modeling. 
\begin{figure}[!htbp]
\centering
\includegraphics[width=0.7\linewidth]{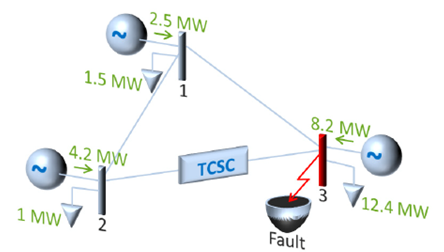}
\caption{Fault in a  BPS with small inertia wind power plant}
\label{fig:milos}
\end{figure}

\begin{figure}[!htbp]
\centering
\includegraphics[width=0.95\linewidth]{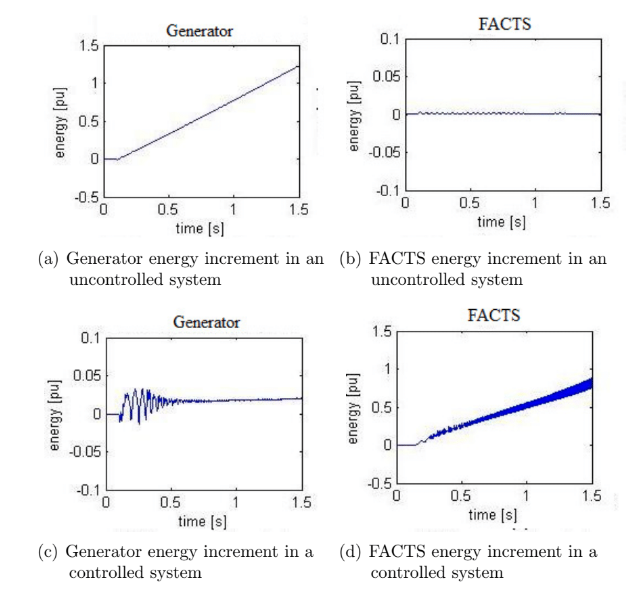}
\caption{Dependence of electro-magnetic dynamics during faults on FACTS control: Increment of accumulated energy cause by a fault - the key role of FACTS control}
\label{fig:faultdyn}
\end{figure}
\begin{figure}[!htbp]
\centering
\includegraphics[width=0.95\linewidth]{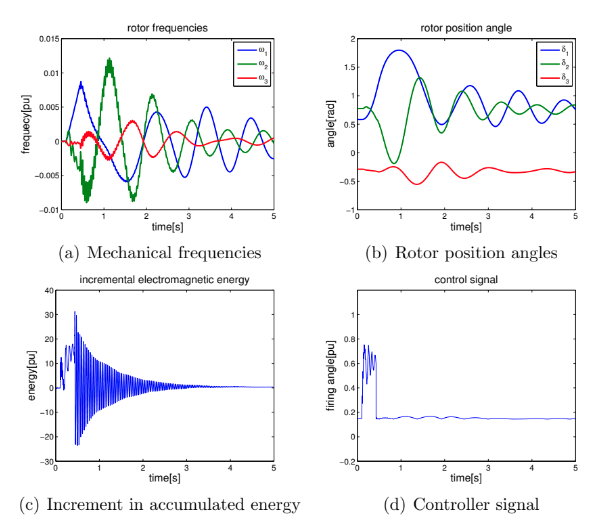}
\caption{Nonlinear control of electromagnetic energy and frequencies}
\label{fig:faultctrl}
\end{figure}
Shown in Figure \ref{fig:milos} is a sketch of a  BPS with one large inertia power plant, small inertia wind power plant sending power to load via TCSC-compensated transmission line \cite{milosfault}. During  short circuit at bus 3, load is lost and electro-mechanical energy in generators begins to increase. Prior to this,  electro-magnetic energy stored in generators increases and it reduces critical clearing time  determined by the electro-mechanical dynamics.  Shown in Figures \ref{fig:faultdyn}  and \ref{fig:faultctrl} is fast accumulated energy in the small inertia generator without TCSC, and it is contrasted with the same electro-magnetic energy  transferred to the TCSC.   This dynamics and the effect of fast controller cannot be modeled unless dynamics of wires and   TCSC  are accounted for. At the same time, it is possible to have a lumped-parameter TVP-based model for designing  effective control logic without requiring excessively detailed EMT  modeling.  Later in Section \ref{sec:control} we  describe energy control design used here. The purpose of showing this example is to demonstrate that during fast sudden changes it becomes necessary to model  electro-magnetic dynamics  and not just the electro-mechanical  dynamics. 

\begin{figure}[!htbp]
\centering
\includegraphics[width=2.5in]{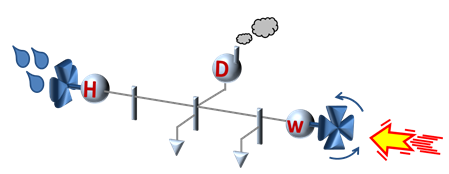}
\caption{Small electric power grid in Flores}
\label{fig:flores}
\end{figure}
Shown in Figure \ref{fig:flores} is a sketch of a small Flores power system in Azores Islands, Portugal \cite{azores}.  
\begin{figure*}
\centering
\includegraphics[width=0.9\linewidth]{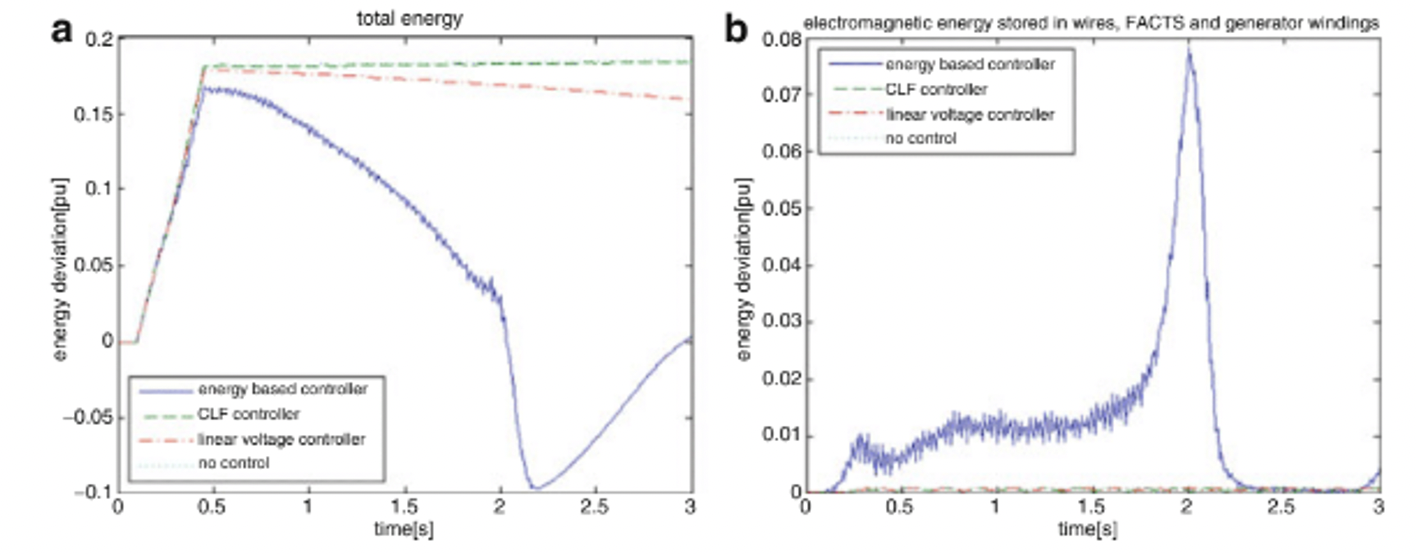}
\caption{SVC control of short wind gust in Flores electric power grid:
(a) Total accumulated energy and (b) total accumulated electromagnetic energy in a
system controlled by different controller}
\label{fig:svcgust_2}
\end{figure*}
\begin{figure}[!htbp]
\centering
\includegraphics[width=0.7\linewidth]{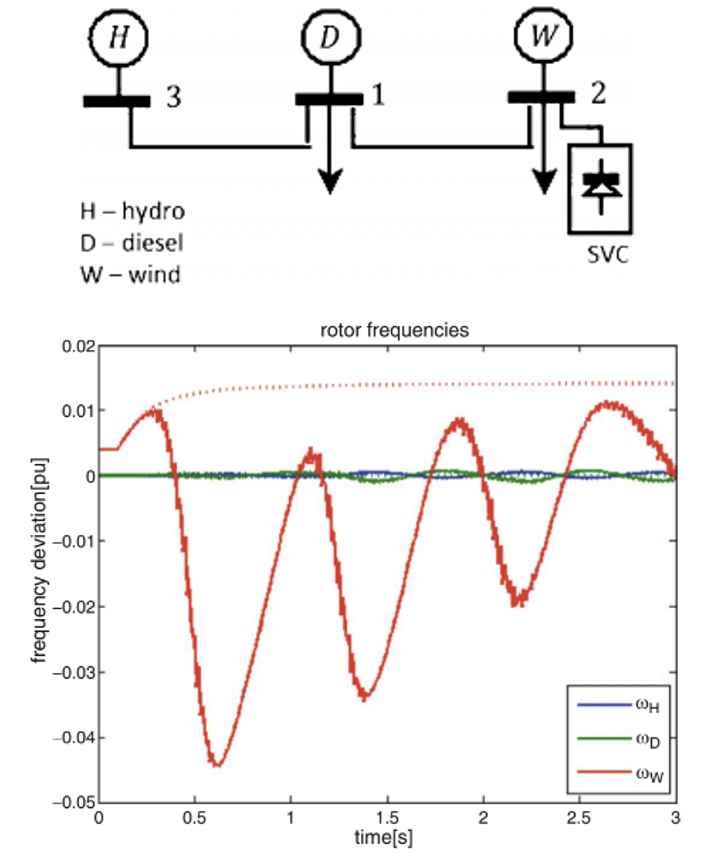}
\caption{SVC control of short wind gust in Flores electric power grid:
Mechanical frequency of all generators in the system during a short-term high-
magnitude wind perturbation: (a)
dashed (without control on the SVC), (b)
solid (with control on the SVC)}
\label{fig:svcgust_1}
\end{figure}
\begin{figure}[!htbp]
\centering
\includegraphics[width=0.8\linewidth]{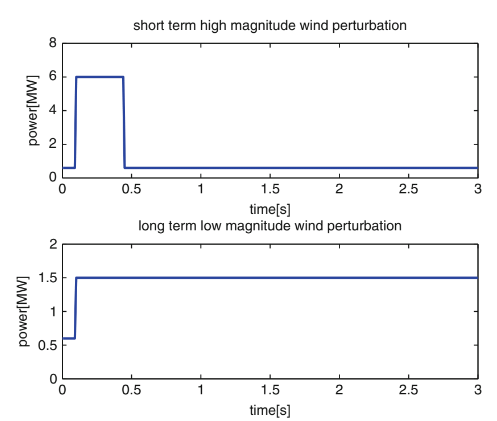}
\caption{Long and short term wind gust in Flores electric power grid}
\label{fig:windgust}
\end{figure}
An SVC can absorb  a sudden power burst by storing electro-magnetic energy fast, as shown in Figure \ref{fig:svcgust_2} and \ref{fig:svcgust_1}, for a sudden wind gust. 
A typical wind gust is shown in Figure \ref{fig:windgust} of short and long duration, respectively.  

\begin{figure}[!htbp]
\centering
\includegraphics [width = 1.0\linewidth]{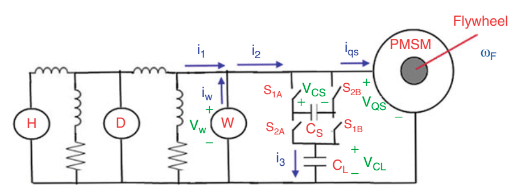}
\caption{Full diagram connecting the flywheel to Flores}
\label{fig:flywheel_2}
\end{figure}

\begin{figure}[!htbp]
\centering
\includegraphics [width = 1.0\linewidth]{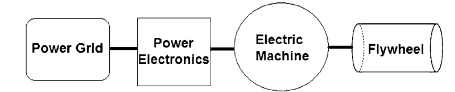}
\caption{Interface between the power grid and the flywheel}
\label{fig:flywheel_1}
\end{figure}

\begin{figure}[!htbp]
\centering
\includegraphics [width = 0.8\linewidth]{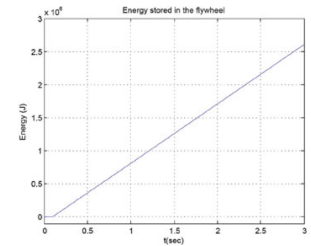}
\caption{Flywheel control of long wind gust in Flores  electric power grid}
\label{fig:flywheel_3}
\end{figure}

Similarly, consider a flywheel storage system connected to Flores island shown in Figure \ref{fig:flywheel_2} (conceptual block diagram shown in Figure \ref{fig:flywheel_1}). The flywheel can compensate for a longer-duration wind gust as shown in Figure \ref{fig:flywheel_3}.

The  TVP-based energy  models can generally be used to design energy  controllers which counteract  oscillations caused by such sudden  disturbances, without requiring EMT modeling. Given inherent multi-layered energy modeling in terms of interaction variables which represent the dynamics of electromagnetic interactions in addition to the dynamics of  electro-mechanical  interactions described in Section \ref{sec:smallfreq} earlier in this paper,  it becomes possible to control CISS problems even in large-scale  changing electric power systems. 

\section{Unifying structure-preserving multi-layered energy modeling}
\label{sec:general}

\begin{figure}[!t]
\centering
\includegraphics[width=0.9\linewidth]{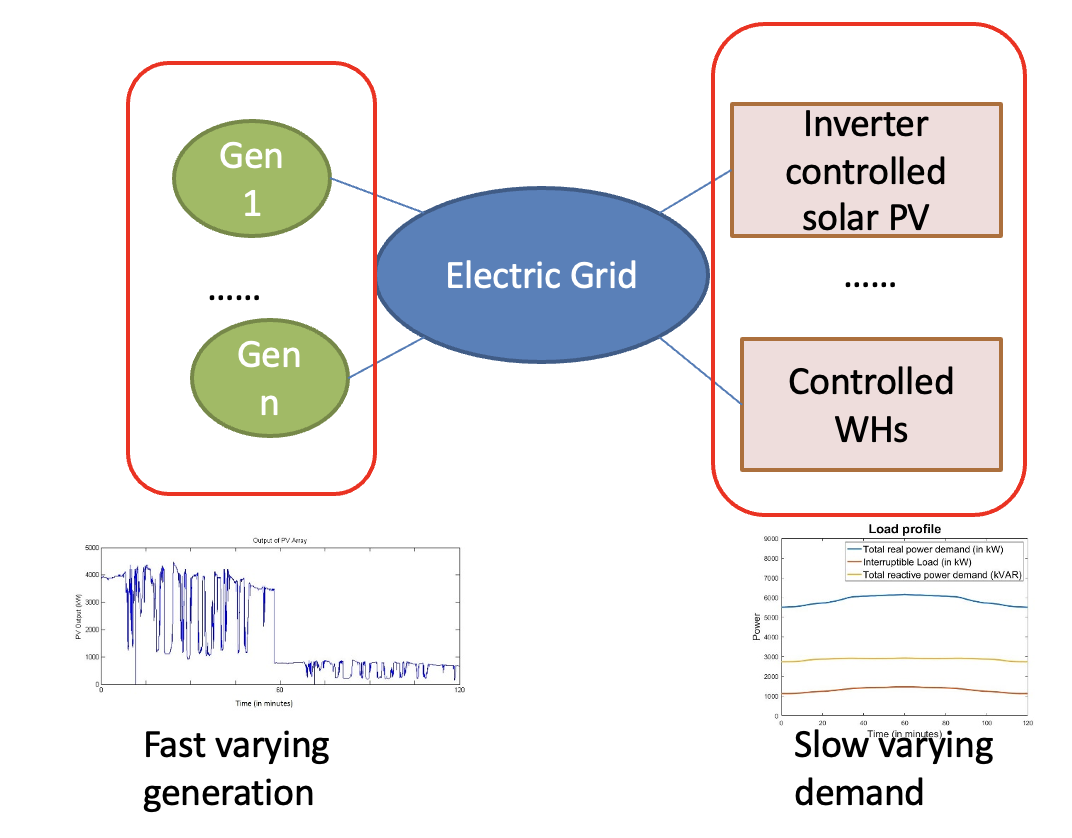}
\caption{IBRs interconnected  to a legacy BPS; fast IBR generation serving slow system load}
\label{fig:3intvar}
\end{figure}

\begin{figure}[!t]
\centering
\includegraphics[width=0.9\linewidth]{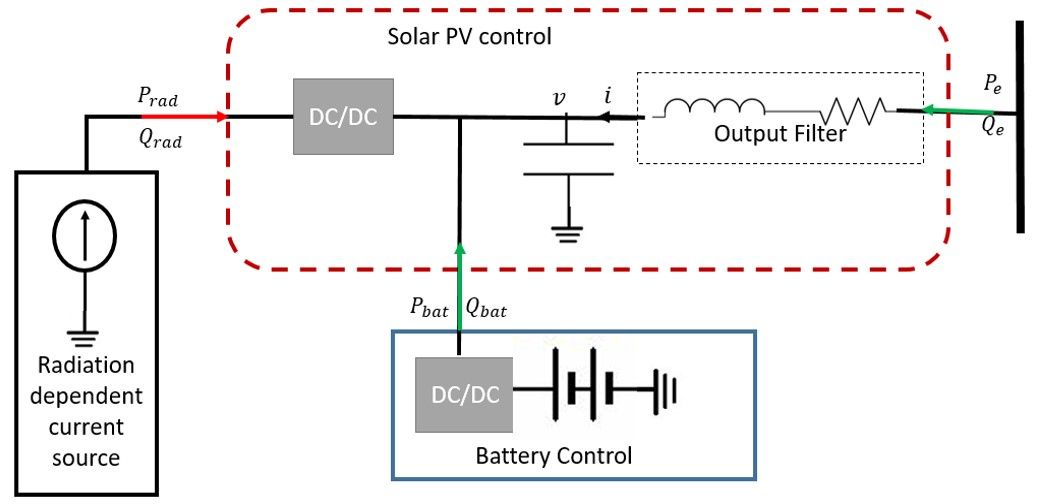}
\caption{Internal design of a typical solar PV with a battery}
\label{fig:solar}
\end{figure}

Shown in Figure \ref{fig:3intvar} is a sketch of legacy system with the  interconnected IBRs.The legacy system comprises conventional power plants serving slowly varying system  load. The IBRs, solar in this example system, send highly fluctuating power injections and are equipped with  fast power-electronically switching inverters.  Shown in Figure \ref{fig:solar}  is internal design of a typical solar PV with small battery. Inverter controllers are not standardised, but for the purposes of this paper they can be thought of black boxes in which some energy is stored, and that they interact with the rest of the system  by balancing energy over certain time,  by balancing instantaneous power at the interconnections according to the first law of thermodynamics, conservation of  energy. However, since the rates of change of instantaneous power are vastly different,  it becomes necessary to also balance power at the right rates to avoid acceleration and related oscillations. 
It was recently proposed to generalize the interaction variable previously used for assessing and controlling  electro-mechanical  dynamics, as described in the section above. 

Such  generalized interaction variable   is defined as \cite{arc,cdc2018}. 
\begin{equation}
\dot z^{out}_i =[P^{out}_i ~~\dot Q^{out}_i]^T
\label{eqn:intgen}
\end{equation}
It  has the same  structural properties as the interaction variable characterizing electro-mechanical  dynamics introduced above in Eqn. (\ref{eqn:intvari}), except that it depends on both its states and rates of change of internal states.  As such,  it is instrumental in modeling interactions  between different modules without making decoupling assumptions  or having to linearize  nonlinear  models.  In particular,  it was shown that the dynamics of aggregate variables, both energy  $E_i$ and its rate of change $p_i= \frac{dE_i}{dt}$ take on a general technology-agnostic form
\begin{eqnarray}
    \dot E_i & = & -\frac{E_i}{\tau_i} + P^{out}_i+ P^{u}_i  = p_i\label{eqn:edyn}\\
    \dot p_i & = & 4 E_{t,i} - \dot{Q}^{out}_i - \dot{Q}^{u}_i
    \label{eqn:pdyn}
\end{eqnarray}
The interaction variables must obey both conservation of instantaneous power and the rate of change of generalized instantaneous power. Basically,  both instantaneous power out of module $i$  and rate of change of instantaneous reactive power should balance with the net  instantaneous power and the net  rate of change of  instantaneous reactive power  being sent from the neighbouring  modules, respectively, as follows
\begin{eqnarray}
P^{out}_i & = &  P^{in}_j\label{eqn:powerbal}\\
\dot Q^{out}_i & = & \dot Q^{in}_j
\label{eqn:qdotbal}
\end{eqnarray}

The aggregate interactive model defining  rate of change of energy and rate of change of reactive power  is given in Eqn.  (\ref{eqn:edyn}) and (\ref{eqn:pdyn}) and is  subject to  conservation laws of  rate of change of  interaction variable defined in Eqn. (Eqn. \ref{eqn:intgen}) as in Eqn. (\ref{eqn:powerbal}) and Eqn. (\ref{eqn:qdotbal}).

\subsection{An example of structure preserving solar PV serving constant power load}
\begin{figure}[!t]
\centering
\includegraphics[width=2.5in]{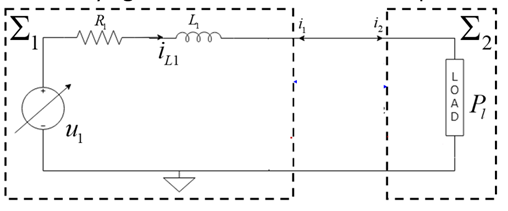}
\caption{Sketch of an inverter controlled solar battery serving power load}
\label{fig:RLckt}
\end{figure}

\begin{figure}[!htbp]
\begin{center}
\includegraphics[width=1.0\linewidth]{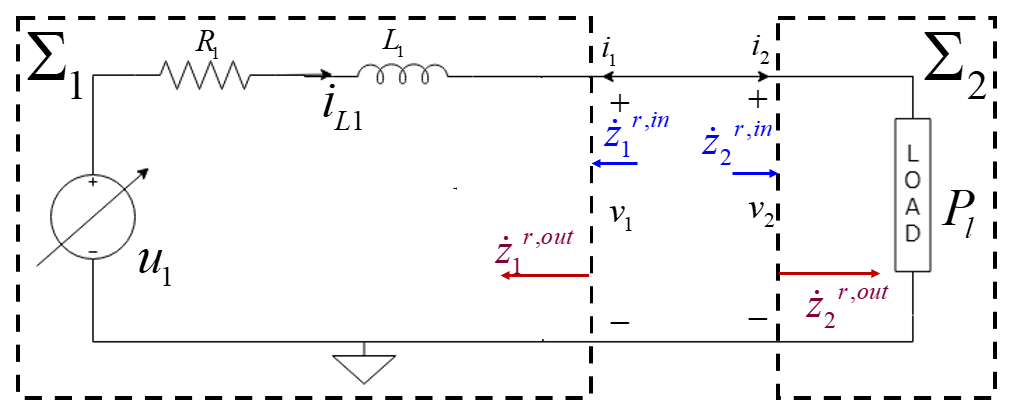}
\caption{Interconnected system comprising interactions between RLC circuit with controllable voltage source ($\Sigma_1$) and a load with given power specifications ($\Sigma_2$)}
\label{fig:RLCCase}
\end{center}
\end{figure}

Shown in Figure \ref{fig:RLckt} is a conceptual sketch of solar PV with a battery \cite{cdc2021}. Internal states are  inductor current and capacitor voltage and inverter-controlled solar source is shown as a controllable DC source. 
There are several different ways of controlling this source, ranging from two-loop  proportional voltage-current controller;  or  PID controller; or energy based control which stabilizes  aggregate variables by aligning the interaction variables between the solar PV and power load. 
In Section \ref{sec:control} we will describe in detail the effects  of primary controllers on closed loop dynamics of inverter-controlled solar PV with a battery. 
For purposes of illustrating further  energy modeling and control we consider without loss of generality   its representation as a voltage controlled source connected via RL circuit,  with the objective  of serving   constant  power load as shown in Fig.  \ref{fig:RLCCase}.
This circuit can be thought of as comprising two interacting modules $\Sigma_1$ and $\Sigma_2$ . The  dynamics of internal state variable, inductor current, takes on the form 
\begin{equation}
    \frac{di_{L1}}{dt} = -\frac{R_1}{L_1} i_{L1} + \frac{1}{L_1}u_1 - \frac{1}{L_1} v_1
\end{equation}
Its energy model takes on the form
\begin{eqnarray}
    \dot E_1 & =  & -\frac{E_1}{\tau_1} + P^{out}_1 + P^{u}_1
    \label{eqn:endynPVE}\\
    \dot p_1  &= &  4E_{t,1} -\dot Q^{out}_1 + \dot Q^{u}_1
    \label{eqn:endynPVp}
\end{eqnarray}
where $E_1 = \frac{1}{2} L i_{L1}^2$, $\frac {E_1}{\tau_1} = Ri_{L1}^2$, $P^{u}_1$ and $\dot{Q}^{u}_1$ are interaction variables between the controllable source and the RL circuit within the module $\Sigma_1$. 
The term $E_{t,1} = \frac{1}{2} L (\frac{di}{dt})^2$ is the stored energy in tangent space.    
The aggregate variables of  this module are $x_{z,1} = [E_1~~p_1]^T$, and  their dynamics takes on the form
\begin{equation}
    \dot x_{z,1} = A_{z,1} x_{z,1} + B_{z} u_{z,1} + B_{t} E_{t,1} +B_{z} \dot z_{1}^{out}
    \label{eqn:aggPVdyn}
\end{equation}
Here, $B_{t} = \left[0, 4\right]^T,  B_z = \left[1, -1\right]^T$ for any component and matrix $A_{z,1} = \left[ {\begin{array}{*{20}{c}}
			{ - 1/{\tau _1}}&0\\
			0&0
			\end{array}} \right]$
			depend only on the time constant $\tau_1$.
			
 Notably, when aggregate control is selected to be $u_{z,1} =[P^{u}_1 ~~\dot{Q}^{u}_1]^T$, it can be designed to 
   align the  rate of change of interaction variable $ \dot z_1^{out}$ and  rate of change of interaction variable  $ \dot z_2^{out}$ from $\Sigma_2$. The closed loop aggregate dynamics is linear and can be shaped  in a provable manner.  This control design is illustrated next.  
\section{Unifying energy control of interaction dynamics
}\label{sec:control}
It follows from the above derivations that  a technology-agnostic energy controller  which aligns rate of change of interaction variables so that the conservation of power and rate of change of reactive power are observed according to Eqns. (\ref{eqn:powerbal}) and (\ref{eqn:qdotbal}). This is achieved by having the control in energy space $u_{z,1}$ respond to the the output variable that takes the form
\begin{equation}
y_{z,1} = C_{t,1} E_{t,1} + C_{z,1} x_{z,1} + D_{z,1} (\dot{z}^{out}_{1} - \dot z_{1}^{ref})
\label{eqn:aggout}  
\end{equation}
where $\dot z_{1}^{ref}= \dot z_{2}^{out}$.  The  energy control can be designed to both cancel error dynamics and the term $E_{t,1}$. Alternatively, the term $E_{t,1}$ can be upper-bounded and the sliding mode control implementation can be done \cite{cdc2021}. 

This technology-agnostic control $u_{z,1}$ needs to be mapped into physically-implementable control signal. In the case of $\Sigma_1$ the physical control signal is $u_1=v_1$. The implementation of this physical control signal is typically power-electronically controlled switch. This map is non-unique and one example of mapping is shown in Eqn. (\ref{Eqn:DynCtrl}) resulting in a dynamical control. 
\begin{equation}
    \frac{du_1}{dt} = \frac{u_1}{i_{L1}} \frac{di_{L1}}{dt} - \frac{\dot{Q}_1^u}{i_{L1}}
    \label{Eqn:DynCtrl}
\end{equation}

\subsection{Dependence of system response on  controller logic  when serving constant power load}
\begin{figure*}
	\centering
	\subfigure[PI two loop controller resulting in voltage collapse]{\includegraphics[width=0.8\linewidth]{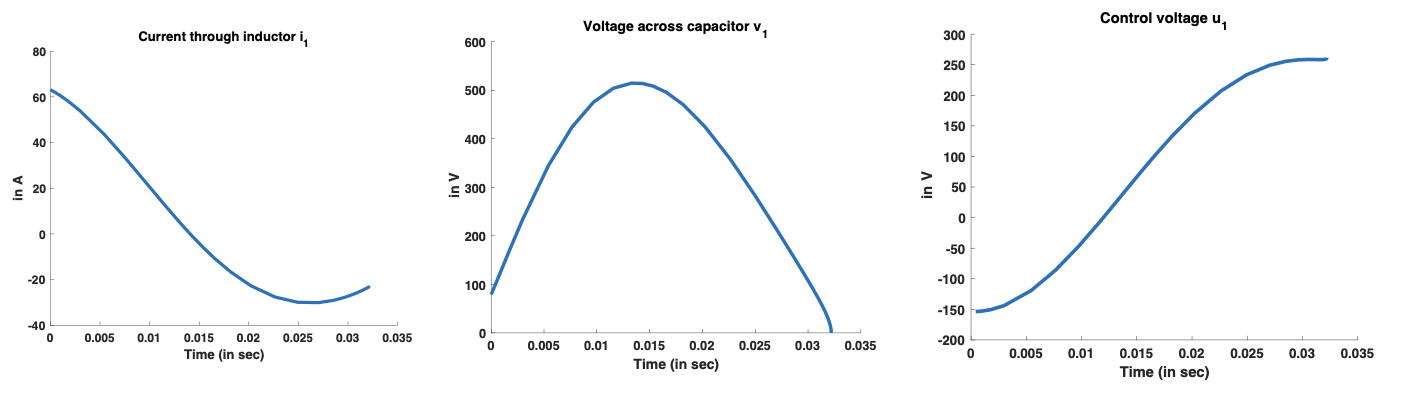}\label{fig:twoloops1}}
	\subfigure[PD controller resulting in stable response]{\includegraphics[width=0.8\linewidth]{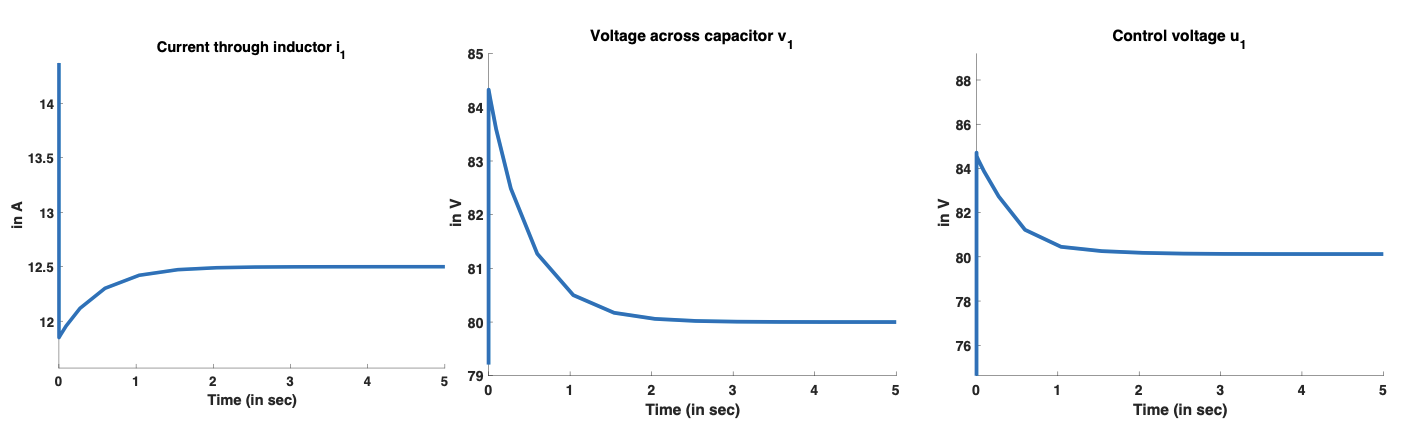}\label{fig:twoloops2}}
	\caption{Closed loop response to step power load  change}. 
	\label{fig:twoloops}
\end{figure*}

As one attempts to control systems with IBRs, it is necessary to assess performance of today's state-of-the-art controllers with the energy  controllers  described in this paper.  The actual expression for energy control differs depending on whether the objective is to:
\begin{itemize}
    \item Single timescale energy control: Align power controlled  by the IBR after its stored energy settles. 
    \item Two timescale energy control: Align both instantaneous power and  rate of change of instantaneous reactive power controlled by the IBR and required by the load. 
\end{itemize}
To illustrate this, consider without loss of generality  a step power change required by the load from the IBR sketched in Figure \ref{fig:RLckt}. This scenario is rather difficult because constant power load is known to create negative incremental resistance problem which inherently leads to voltage collapse and/or other types of unacceptable dynamics in IBRs controlled by conventional controllers. Shown in Figure \ref{fig:twoloops1} is the  response of such controller. It can be seen that  typically-used two loops control will have problems serving constant power load as shown in Figure \ref{fig:twoloops1}, voltage collapsing after several seconds.  To the contrary, the response of  a conventional proportional-derivative  (PD) controller remain stable, as shown in the plots in Figure \ref{fig:twoloops2}. This plot highlight the key relevance of dynamic controllers. 
\begin{figure*}
\centering
\includegraphics[width= 0.8\linewidth]{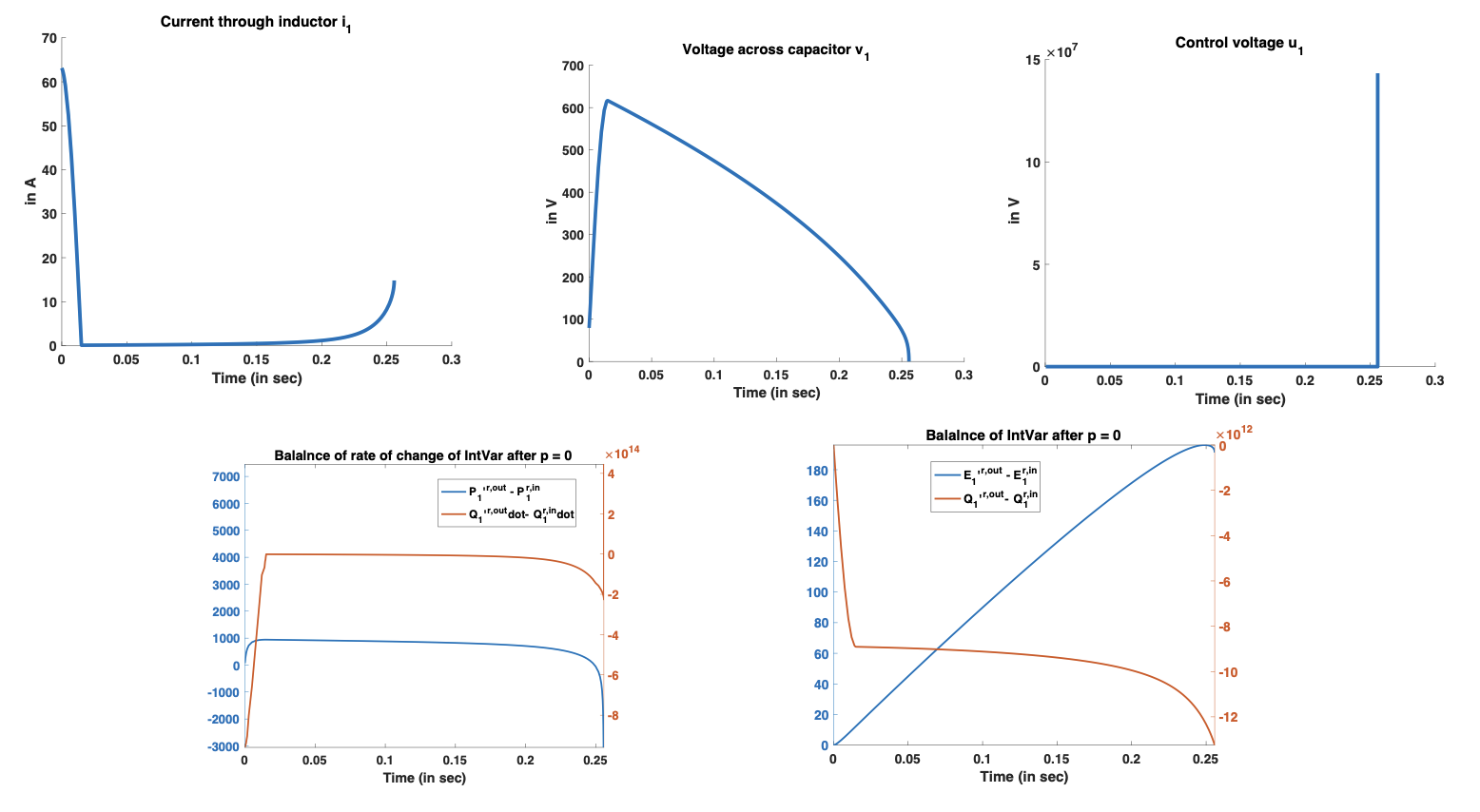}
\caption{Closed loop response to step power load  change: 
Single timescale nonlinear energy controller;  
Upper plots show the state trajectories. Lower plots show the evolution of the difference in the outgoing and incoming interaction variables.
}
\label{fig:singlescale}
\end{figure*}
\begin{figure*}
\centering
\includegraphics[width=0.8\linewidth]{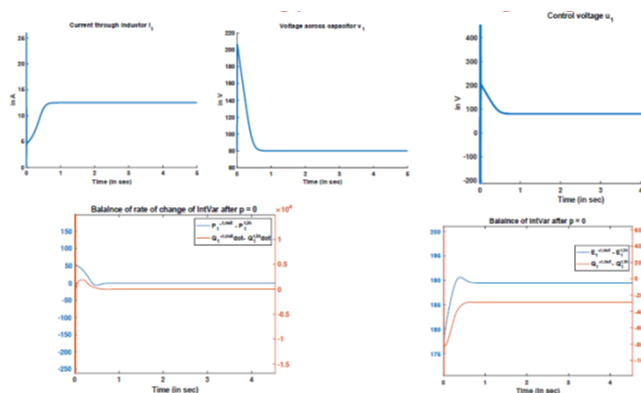}
\caption{Closed loop response to step power load  change: 
Two-timescale nonlinear energy controller; Lower plots show the evolution of the difference in the outgoing and incoming interaction variables.}
\label{fig:twoscales}
\end{figure*}

\begin{figure*}
\centering
\includegraphics[width=0.8\linewidth]{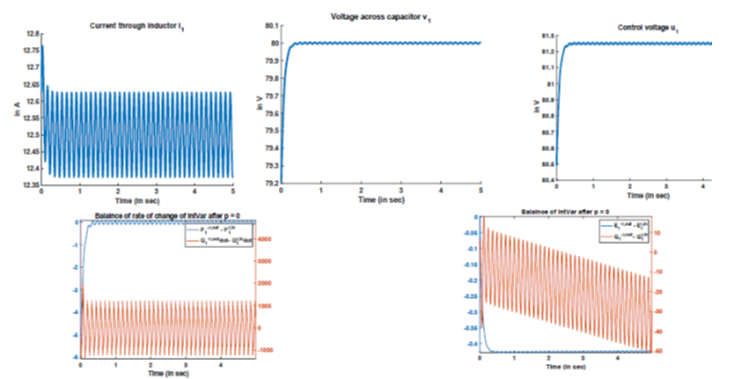}
\caption{Closed loop response to small intermittent fluctuations: Conventional feedback control}
\label{fig:convfluct}
\end{figure*}
\begin{figure*}
\centering
\includegraphics[width=0.8\linewidth]{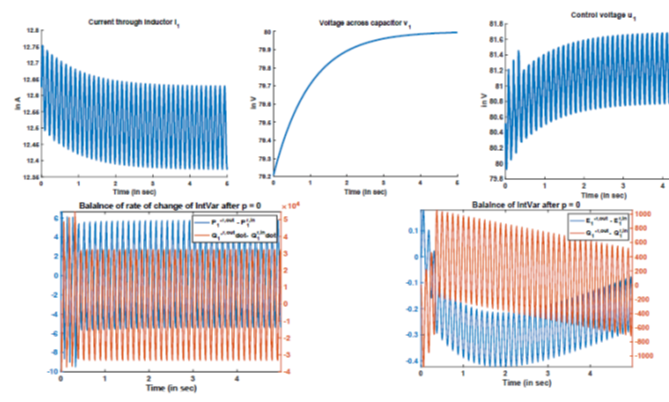}
\caption{Closed loop response to small intermittent fluctuations: Constant gain derivative control responding to voltage}
\label{fig:derv}
\end{figure*}
\begin{figure*}
\centering
\includegraphics[width=0.8\linewidth]{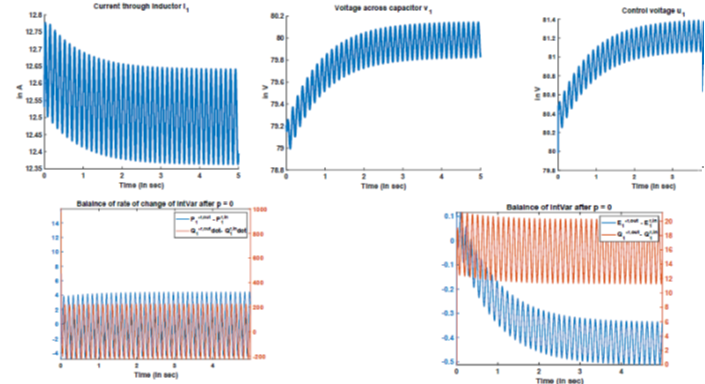}
\caption{Closed loop response to small intermittent fluctuations: 
Single timescale nonlinear energy controller; Lower plots show the evolution of the difference in the outgoing and incoming interaction variables.
}
\label{fig:lowgain}
\end{figure*}
\begin{figure*}
\centering
\includegraphics[width=0.8\linewidth]{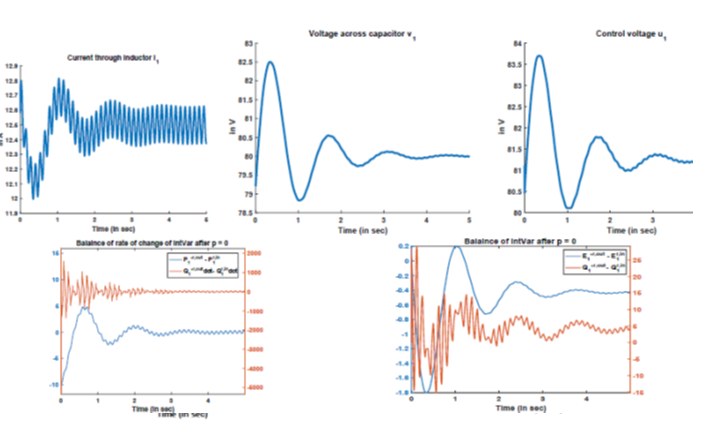}
\caption{Closed loop response to small intermittent fluctuations: 
Two-timescale nonlinear energy controller; Lower plots show the evolution of the difference in the outgoing and incoming interaction variables.
}
\label{fig:highgain}
\end{figure*}
Shown next in Figure \ref{fig:singlescale}  is the closed loop dynamic response for the same scenario with the energy controllers when both electro-mechanical and electro-magnetic dynamics are controlled as a single time scale process, and when the only  control objective is to  alignment of instantaneous power after energy  dynamics settles ($\frac{dE}{dt} =0$). It can be seen that even  this nonlinear controller is not fast enough to stabilize system response and to avoid voltage collapse. 

However, when the energy controller is designed with the objective to align both instantaneous power and rate of change of instantaneous reactive power at the interfaces of IBR and constant power load, and time scale separation between these processes is carefully accounted for by using slow reduced order model to stabilise electro-mechanical  dynamics and fast reduced order model is used to stabilise fast reactive power rate of change,  shown in Figure \ref{fig:twoscales} is closed loop stable response. 

\subsection{Dependence of system response on  controller logic  when serving intermittent power load}
A similar comparison of the effects of iBR controllers on closed loop response can be shown in response to fluctuating power disturbances seen by the IBR.  
Shown in Figure \ref{fig:convfluct} is the closed loop response to small intermittent fluctuations.  It can be seen that conventional feedback is too slow to cancel the fast imbalances, resulting in small persistent oscillations in voltage. 
The response obtained with dynamical PD controller cancels the persistent oscillations, but results in slower response as shown in Fig. \ref{fig:derv}.

The response obtained with single timescale energy control is shown in Fig. \ref{fig:lowgain}. It can be seen that even this nonlinear controller is not able to suppress fast imbalances. Finally, the responses obtained with two timescale energy control is shown in Fig. \ref{fig:highgain}, which results in perfect voltage regulation. We further see that the instantaneous and reactive power imbalances settle at zero. 
%
%
%
\section{Concluding recommendations: Structure-preserving standards/protocols for reliable and efficient multi-energy systems}
\label{sec:standards}
In closing, this paper takes a broad look at the state-of-the-art hierarchical modeling and control for bulk power systems. It identifies several hidden roadblocks to using  this paradigm in support of large-scale decarbonization. Fundamental challenges are described first.  It is  proposed next to utilize structure-preserving modeling of complex power grids which has been introduced by our  many collaborators and former doctoral students both at Carnegie Mellon University and, more recently at M.I.T. These concepts are  explained in the context of emerging low frequency oscillations,  using the notion of interaction variable.  Then a recent generalization of interaction variables is introduced and it is suggested that energy modeling which captures both instantaneous power and instantaneous rate of change of reactive power  be considered as the basis for multi-layered modeling of the changing electric energy systems. An aggregate model of technology-agnostic module is derived in terms of stored energy and its rate of change, and it is used for nonlinear energy control. A proof-of-concept energy control which avoids voltage collapse and  significantly reduces otherwise persistent oscillations  created by intermittent  IBRs is presented using a DC voltage controlled source which supplies time varying power.    Based on these findings, further research is needed toward a modular feed-forward  distributed information exchange  in support  of self-adaptation of subsystems to ensure feasible interconnected system. 

We propose the interaction-based energy modeling to be used as the basis for structure preserving interacting  standards/protocols that ensure stable  operations in the changing systems comprising multiple layers of intelligent Balancing Authorities (iBAs) with well-defined interfaces to the rest of the system. We suggest   three steps toward generalising today's area control error (ACE) as the basis for  reliable and efficient  emerging  systems operation.  They are as follows:
\begin{itemize}
\item Generalise today's coarse balancing authorities into intelligent Balancing Authorities (iBAs).
\item  Enhance today's AGC standard  Area control error (ACE) into generalized interaction variables as the measure of any iBAs performance. 
\item Work toward next generation end-to-end SCADA supported by the information exchange proposed.  
\end{itemize}
These three steps form the foundations of interactive operations in future energy systems so that they can be modelled, controlled and operated without excessive complexity. 

\section*{Acknowledgement}
The  first author greatly appreciates partial  support from the  NSF EAGER  Project   number 2002570
entitled  ``EAGER: Fundamentals of Modeling and Control for the Evolving Electric Power System Architectures". 
The first author fondly recalls Dr Narain Hingorani's comments at the CURENT lunch meeting, in which stated that  ``it is all about energy dynamics" and not about ``low inertia". This comment has helped us pursue what is presented in this paper.
The first   author is also very grateful for many years of  collaborations with students and colleagues, those whose papers are referred to and all others,  who have helped with concepts presented here.  

\IEEEtriggeratref{12}


\bibliographystyle{IEEEtran}
\bibliography{references}
%

\end{document}